\documentclass[preprint,showpacs,preprintnumbers,amsmath,amssymb]{revtex4}

\usepackage{graphicx}
\usepackage{dcolumn}
\usepackage{bm}

\newcommand{\be}{\begin{equation}}
\newcommand{\ee}{\end{equation}}
\newcommand{\ben}{\begin{equation*}}
\newcommand{\een}{\end{equation*}}
\newcommand{\bea}{\begin{eqnarray}}
\newcommand{\eea}{\end{eqnarray}}
\newcommand{\bean}{\begin{eqnarray*}}
\newcommand{\eean}{\end{eqnarray*}}
\newcommand\re[1]{(\ref{#1})}

\newcommand{\eqnsection}{
\renewcommand{\theequation}{{\thesection.\arabic{equation}}}
\makeatletter \csname @addtoreset\endcsname{equation}{section}
\makeatother} \eqnsection

\begin{document}
\title{Unique additive information measures -
Boltzmann-Gibbs-Shannon, Fisher and beyond}
\author{P. V\'an}
\address{Budapest University of Technology and Economics\\
Department of Chemical Physics\\
1521 Budapest, Budafoki \'ut 8.}

\email{vpet@eik.bme.hu}

\pacs{05., 02.50.-r, 89.70.+c}

\date{\today}

\begin{abstract}

It is proved that the only additive and isotropic information
measure that can depend on the probability distribution and also on
its first derivative is a linear combination of the
Boltzmann-Gibbs-Shannon and Fisher information measures. Power law
equilibrium distributions are found as a result of the interaction
of the two terms. The case of second order derivative dependence is
investigated and a corresponding additive information measure is
given.
\end{abstract}

\maketitle

\section{Introduction}

In 1957 in his famous paper on information theory and statistical
mechanics, Jaynes suggested to look statistical mechanics as a form
of statistical inference. He argued that the usual rules of
statistical mechanics are justified independently of experimental
verification and additional physical arguments, because "they still
represent the best estimates that could have been made on the basis
of information available" \cite{Jay57a}. He recognized the physical
importance of the train of thought of Shannon, where it was proved
that the logarithmic form of the information measure is a
consequence of some simple properties that any information measure
should have. Based on this observation, he suggested to start
statistical physics from a maximum entropy principle.

Dealing with a discrete probability space where a variable $x$ can
assume the discrete values $(x_1, ..., x_n)$ with the corresponding
probabilities $(p_1, ..., p_n)$ there exist a function
$H(p_1,...,p_n)$, which is
\begin{itemize}
 \item continuous,
 \item with equal probabilities is monotone increasing with $n$,
 \item satisfies the composition law.
\end{itemize}
These conditions are those that we would expect from a measure of
information. With the above properties the function $H(p_1,...,p_n)
= -k \sum_{i=1}^n p_i \ln p_i$ is {\em unique} up to the positive
multiplier $k$.

Later these conditions were investigated, generalized and clarified
extensively both from mathematical and from physical points of view.
It turned out that continuity is not so important. Monotonicity can
be replaced by {\em concavity} and determines the sign of the
function only. The most important assumption from a physical point
of view is the composition law. Later the composition law was
reformulated in a more convenient way as {\em additivity}. R\'enyi
recognized that the only possibility to find a different additive
measure of information is to generalize also the method of averaging
\cite{Ren61a}.

The uniqueness is the key feature that connect the information
measure to the physical entropy and ensures that the consequences of
the macroscopic Second Law (first of all the existence of the
universal, absolute temperature) are valid for the quantities of the
statistical physical approach. Because of the uniqueness, the
arising statistics is the same, independently of the microscopic
dynamics.

As additivity gives the connection between the statistical and
thermodynamic theories, its investigation is particularly
interesting if one would like to explain phenomena that is seemingly
out of the framework of traditional methods of statistical physics
\cite{Tsa88a,Kan01a,Coh02a,BecCoh03a}.

In all previously mentioned researches it was assumed explicitly
that the entropy is a {\em local} function of the variables. In this
paper we weaken this assumption and look for additive entropy
functions interpreted on a continuous state space, that depend not
only on the probability distribution but also on the derivatives of
the mentioned probability distribution. In the following we will
prove generalizations of the statement of Shannon for derivative
dependent information measures. In the next section we prove, that
the unique additive, continuously differentiable entropy/information
measure that depends only on the first derivatives of the
probability distribution is a linear combination of the
Boltzmann-Gibbs-Shannon and the Fisher information measures. In the
third section we investigate the physical meaning of the constructed
unique information measure. Then we construct an additive
information measure that contains second order derivatives. Finally
there are some conclusions.

\section{Weakly nonlocal information measures: first order}

Let us consider an $n$ dimensional continuous probability space
$X\subset \mathbb{R}^n$, where the probability measure can be given
by a continuously differentiable nonnegative function, the
probability density $f:\mathbb{R}^n \rightarrowtail \mathbb{R}^+$,
which is normalized, \be \int_X f(x) dx =1. \label{normc}\ee

A first order weakly nonlocal information measure is a function
$s(f, Df)$ of the probability density $f$ and its derivative $Df$
with some expected properties. An information measure is positive,
increases with increasing uncertainty, and is additive for
independent sources of uncertainty. In case of derivative dependent
information measures it is convenient to require the isotropy of
$s$, too. These conditions can be formulated as follows
\begin{enumerate}
\item {\em Isotropy}. An isotropic function $s$ of $f$ and $Df$
has the following form
\begin{equation}
s(f,Df) = \hat{s}(f,(Df)^2). \label{isotrop}\end{equation}

\item {\em Additivity}. For the sake of simplicity we restrict
ourself for two independent distribution functions $f_1({\bf x}_1)$
and $f_2({\bf x}_2)$ defined on $X=X_1 \times X_2 \subset
\mathbb{R}^{n_1}\times \mathbb{R}^{n_2}, \quad n_1+n_2=n$. The
generalization to finite number of distributions is straightforward.
Then additivity requires
\begin{equation}
s_n(f_1f_2,D(f_1f_2)) = s_{n_1}(f_1,Df_1)+s_{n_2}(f_2,Df_2),
\label{additiv}\end{equation}

\noindent where the subscripts denote the different dimensions of
the domains.
\end{enumerate}

Without isotropy additivity cannot be formulated easily because the
domain of the function $\hat{s}$ is the same on both sides of the
above formula. Although most probability distributions in physics
are defined on spaces that are highly anisotropic, here we restrict
ourselves to isotropic information measures on isotropic state
spaces and leave that problem for further investigations.
Fortunately, in the simplest situation, when the state space is the
Descartes product of isotropic subspaces (position-momentum) one can
keep the simple formulation of additivity with some straightforward
assumption.

For independent probability distributions the unified probability
density $f({\bf x}_1,{\bf x}_2)$ is the product of the probability
densities $f_1({\bf x}_1)$ and $f_2({\bf x}_2)$. Thus, we have
$Df({\bf x}_1,{\bf x}_2) = (f_2({\bf x}_2)D_{{\bf x}_1}f_1({\bf
x}_1),$ $f_1({\bf x}_1)D_{{\bf x}_2}f_2({\bf x}_2))$ and — omitting
the variables ${\bf x}_1$ and ${\bf x}_2$ — $(Df)^2 = (f_2 Df_1)^2 +
(f_1 Df_2)^2$. As a consequence, for isotropic information measures
the additivity requirement can be written as
\begin{equation}
\hat{s}(f_1f_2, \: (f_2 Df_1)^2 +
    (f_1 Df_2)^2) = \hat{s}(f_1,(Df_1)^2) +
    \hat{s}(f_2,(Df_2)^2).
\label{defadd}\end{equation}

Differentiating the above equality by $(Df_1)^2$ and $(Df_2)^2$,
respectively we have that
\begin{eqnarray*}
f_2^2 \partial_2 \hat{s}(f_1f_2,\: (f_2Df_1)^2 + (f_1Df_2)^2)
    &=& \partial_2 \hat{s}(f_1,(Df_1)^2), \\
f_1^2 \partial_2\hat{s}(f_1f_2,\: (f_2Df_1)^2 + (f_1Df_2)^2)
    &=& \partial_2\hat{s}(f_2,(Df_2)^2).
\end{eqnarray*}

Here $\partial_2$ denotes the partial derivative of $\hat{s}$ by its
second argument. Therefore
\begin{equation*}
f^2 \partial_2\hat{s}(f,(Df)^2) = -\kappa_1 = const.,
\end{equation*}

\noindent hence
\begin{equation}
\hat{s}(f, (Df)^2) = -\kappa_1\frac{(Df)^2}{f^2} + \tilde{s}(f).
\label{Fisform}\end{equation}

Here $\tilde{s}$ is an arbitrary function (the local part of the
entropy). Repeating the above train of thought with the derivatives
by the first argument of $\hat{s}$, one finds that
\begin{equation*}
f \partial_1\tilde{s}(f) = -\kappa = const.
\end{equation*}

Consequently, $\tilde{s}(f) = -\kappa \ln f + s_0$, where $s_0=0$ by
further applying additivity. Therefore, the most general isotropic
and additive first order weakly nonlocal information measure is
\begin{equation}
s(f, Df) = -\kappa_1 \frac{(Df)^2}{f^2} - \kappa \ln f.
\label{EPIform}\end{equation}

The first term has the form of a Fisher information
\cite{Fis59b,Bor98b} and the second term has the form of a Shannon
information measure. It is clear from the previous calculations that
\re{EPIform} is unique with the above requirements (isotropy and
additivity).

\section{Power law tails in microcanonical and canonical ensembles}

There are several attempts to find the physical significance of
Fisher information (see e.g. \cite{Fri98b,PlaAta05a}). The
observation in the previous section puts these investigations into a
new light. Accepting Jaynes approach in suggesting the central role
of information in statistical physics one should require the
extremum of the two terms together. However, even if we accept the
idea of Jaynes, that the unique information measures are important
from a physical point of view there are several important questions
to be answered. E. g. What is the physics behind the second term?
What could be the value of the constant $\kappa_1$?

Let us consider a classical one dimensional ideal gas where the
Hamiltonian is given as $ H(p) = \frac{p^2}{2m}$, where p is the
momentum and m is the mass of the particles. In this case, according
to the maximum entropy principle one should find the maximum of the
average of the entropy density \re{EPIform} subject to the
constraints of fixed average energy and normality. Therefore we face
to the following variational problem:
\begin{equation}
\int \left( -\kappa_1 \frac{(Df)^2}{f} -
    \kappa f \ln f \right) dp -
    \beta \left(\int f \frac{p^2}{2m} dp - E \right) -
    \alpha \left(\int f dp - 1\right)  = extremum
\end{equation}

In this case the partition function formalism does no help, the
above variational problem leads to the following Euler-Lagrange
equation for $R = \sqrt{f}$:
\begin{equation}
4 \kappa_1 R'' - 2 \kappa R^2 ln R + \frac{\beta}{2 m} p^2 R +
(\alpha - \kappa) R =0. \label{Schreq}\end{equation}

Here the dash denotes the derivation $R'=\frac{{\rm d} R}{{\rm d}
p}$. The corresponding natural boundary condition can be interpreted
as an entropy current $J_s = R'\delta R$ \cite{Mac92b}.

\begin{figure}[ht]
\centering
\includegraphics[height=9cm]{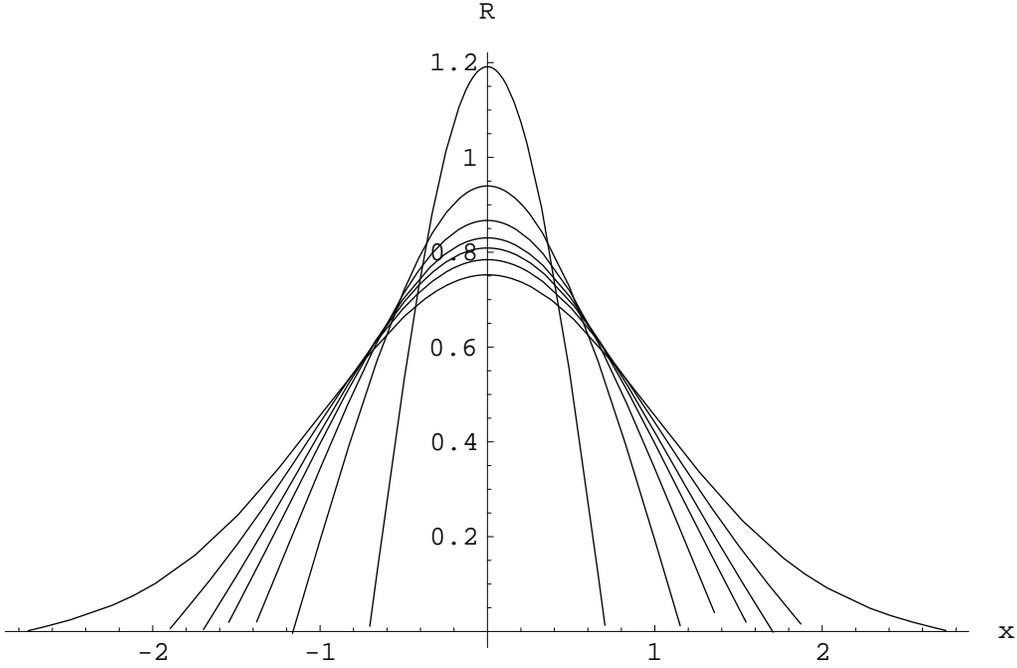}
\caption{The square root of the distribution function f(p) for
different values of  $\Lambda=(1,1.1,1.2,1.3,1.5,2,5)$.}
\label{Fig1}
\end{figure}

For any positive $\kappa_1$ the solutions of the above equation are
different of the classical Maxwell-Boltzmann distribution, but the
properties of the distribution are similar. Let us choose
$\frac{\kappa}{2 \kappa_1} = 0.1$, $\frac{\beta}{8 m \kappa_1} = 1$
and $\frac{\alpha - \kappa}{4 \kappa_1} =: \Lambda$. The symmetric
solutions of the above equation with the condition $R'(0)=0$ have
finite support if $\alpha>0.9897$, the value where we can get back a
Maxwell-Boltzmann like Gaussian distribution. Above that value the
function have power law tail of the form $R_{tail}(p) = D \pm C
p^\gamma$ as one can see from Figure \ref{Fig1}.

We have got a one parameter family of power law tail distributions
parameterized by the Lagrange multiplier $\alpha$ ($\Lambda$). The
parameter can be regarded to different nonzero entropy currents at
the boundary as one can see from the boundary conditions defined
above. The arising family of power law tail distributions are
different form the distributions of nonadditive (nonextensive)
statistics of Tsallis, Beck-Cohen or Kaniadakis
\cite{Tsa88a,BecCoh03a,KanAta04a} and were defined by a unique and
additive information measure.

A related answer to the above questions emerges considering quantum
mechanics as a particular application and focusing on the concavity
properties of the Fisher term. With a suitable reinterpretation of
the terms one can recognize a time independent Schr\"odinger
equation of a harmonic oscillator in \re{Schreq} if $\kappa=0$. The
connection of quantum mechanics to Fisher information was pointed
out by several independent researches \cite{FriAta02a,Reg98a}. We
may determine the $\kappa_1$ constant in these systems. E.g. for a
single particle system one can get, that $\kappa_1 = \hbar^2$.

The understanding the role of Fisher information in quantum
mechanics can give some clues to the further understanding the
physics behind \re{EPIform}. In quantum mechanics only the Fisher
term appears and one should explain the missing
Boltzmann-Gibbs-Shannon term in an information theoretical approach.
In this respect Hall and Regginato argued with strengthening some
basic laws of quantum mechanics and introduced the {\em exact
uncertainty principle} \cite{HalReg02a,HalReg02a1}. V\'an and
F\"ul\"op suggested {\em mass scale invariance}, requiring the
possibility of particle interpretation for the probability
distribution \cite{VanFul03m}.

On the other hand Bialynicki-Birula and Mycielski gives an example
that that the quantum potential could be supplemented by a
Boltzmann-Gibbs-Shannon term \cite{BiaMyc76a}. The solution of the
supplemented Schr\"odinger equation gives non dispersive free
solutions, the so-called "Gaussons", as a result of the additional
logarithmic term.

\section{Weakly nonlocal information measures: second order}

One can ask about properties of information measures depending on
higher order derivatives of the distribution function. Here we
investigate the second order case. The requirements are similar as
previously
\begin{enumerate}
\item {\em Isotropy}. According to the representation theorems of
isotropic functions that depend on a vector ($Df$) and a symmetric
second order tensor ($D^2f$) we can write \cite{Wan70a}:
\begin{multline}
s(f,Df,D^2f) = \\
    \hat{s}(f,\: (D f)^2,\: Df\cdot D^2f\cdot Df,\:
    Df\cdot D^2f \cdot D^2f \cdot Df,... ,
     Df\cdot (D^2f)^{n-1}\cdot Df,\nonumber\\
    Tr (D^2f),\:  Tr(D^2f \cdot D^2f), ...,Tr((D^2f)^n) ).
\label{isotrop2}\end{multline}

\item {\em Additivity}. Here we require that
$$
s(f_1f_2,D(f_1f_2),D^2(f_1f_2)) =
s(f_1,Df_1,D^2f_1)+s(f_2,Df_2,D^2f_2)
$$
\end{enumerate}

As previously, we need to consider isotropy in the formulation of
additivity. Let us observe, that the above form \re{isotrop2} is
restricted very much by the requirement of additivity. Simple
calculations show that the second order nonlocal information measure
is more difficult than the first order one. Its form depend on the
dimension of the phase space. As an example I give the general
additive version of the following isotropic function (this is the
unique general form of weakly nonlocal information measure in three
dimension)
\begin{multline}
s_3(f,Df,D^2f) =
    \hat{s}_3(f,(D f)^2, Df\cdot D^2 f\cdot Df, Df\cdot (D^2 f)^2\cdot Df,
    Tr (D^2f), \\
    Tr(D^2f)^2, Tr(D^2f)^3).
\end{multline}
One can derive that
\begin{multline}
s_3(f,Df,D^2f) =
  -\kappa ln f - \kappa_1 \frac{(D f)^2}{f^2} -
  (\kappa_2+\kappa_5) \frac{(D f)^4}{f^4} +
  (\kappa_3+\kappa_6) \frac{(D f)^6}{f^6} -\\
  \kappa_2 \frac{1}{f^3} Df\cdot D^2 f\cdot Df +
  (2 \kappa_3 + 3\kappa_6) \frac{(D f)^2}{f^5}Df\cdot D^2 f\cdot Df -
  \kappa_3 \frac{1}{f^4} Df\cdot (D^2 f)^2\cdot Df -
  \kappa_4 \frac{1}{f} Tr (D^2f) -\\
  \kappa_5 \frac{1}{f^2} Tr(D^2f)^2 -
  \kappa_6 \frac{1}{f^3} Tr(D^2f)^3.
\label{EPIform1}\end{multline}

The concavity properties of the above function are not
straightforward.

\section{Conclusions}

There are several attempts to understand the reason of the
appearance of Fisher information in different disciplines of physics
\cite{PlaPla95a,Fri90a,FriAta02a,Fri98b}. The usual justification
and interpretation is based on estimation theory. The above proof of
uniqueness explains, why any dynamical background that preserves the
additivity - gives the same Fisher like form, independently of the
estimation theoretical background.

We have seen that for information measures with second order
derivatives the number of additive terms depends on the dimension of
the probability (phase) space. Therefore the physical significance
of information measures containing higher than first order
derivatives is dubious.

There are more questions than answers in this work. However, in the
relationship of thermodynamics and statistical physics the idea of
Jaynes is the key of understanding that deserves further
investigations.

\section{Acknowledgements}

This research was supported by OTKA T034715, T034603 and T048489.
Thank for T. Matolcsi for careful reading of the manuscript.

\end{document}